\documentclass[aps,pre,twocolumn,a4paper,footinbib,longbibliography,showpacs,groupedaddress,floatfix,superscriptaddress]{revtex4-2}%

\usepackage[dvipsnames]{xcolor}
\definecolor{linkcolor}{rgb}{0.3,0.3,1.0} 
\usepackage[pdftex,colorlinks=true, linkcolor= linkcolor, citecolor= linkcolor, urlcolor= linkcolor, hyperindex=true,hyperfigures=true]{hyperref} 

\usepackage{amssymb,amsmath,amsfonts}
\usepackage{graphicx}
\usepackage{float}
\usepackage{bm}
\usepackage{physics}
\usepackage{mathtools}

\usepackage{xcolor}
\definecolor{mygreen}{RGB}{80, 185, 148}
\definecolor{myblue}{RGB}{150, 150, 200}
\definecolor{mygray}{RGB}{180, 180, 180}

\newcommand{\inflow}{\mathcal{G}}

\newcommand{\dotnice}[1]{\overset{\,\bm.}{#1}{\vphantom{#1}}}
\newcommand{\ddotnice}[1]{\overset{\,\bm.\bm.}{#1}{\vphantom{#1}}}

\renewcommand{\vec}[1]{\bm{#1}}

\begin{document}

\title{Balancing information and dissipation with partially observed fluctuating signals} 

\author{Giorgio Nicoletti}
\affiliation{Quantitative Life Sciences section, The Abdus Salam International Center for Theoretical Physics (ICTP), Trieste, Italy}
\author{Ivan Di Terlizzi}
\affiliation{Max Planck Institute for the Physics of Complex Systems, Dresden, Germany}
\affiliation{Ludwig-Maximilians-Universit\"at M\"unchen, Arnold-Sommerfeld-Center for Theoretical Physics, M\"unchen, Germany}
\author{Daniel Maria Busiello}
\affiliation{Max Planck Institute for the Physics of Complex Systems, Dresden, Germany}
\affiliation{Department of Physics and Astronomy ``Galileo Galilei'', University of Padova, Padova, Italy}

\begin{abstract}
Biological systems sense and extract information from fluctuating signals while operating under energetic constraints and limited resolution. We introduce a general chemical model in which a sensor, coupled to a signaling pathway activated by hidden signals, can allosterically tune the production of a readout molecule. We propose viable strategies for the sensor to estimate, and eventually balance, information gathering on the hidden process and the associated dissipative cost relying solely on counting statistics of observed trajectories. We show that these strategies can be successfully implemented to adapt the readout production even with finite-time measurements and limited dynamic resolution, and remain effective in the presence of inhibitory regulatory mechanisms. Our study provides a plausible mechanism to actively balance information and dissipation, paving the way for an implementable design principle underpinning biological and biochemical adaptation.
\end{abstract}

\maketitle

Biological and synthetic systems often rely on noisy fluctuating signals to guide their behavior \cite{bowsher2014environmental,reddy2016learning,balazsi2011cellular,karin2021temporal,nakajima2015biologically}. Cells, for instance, infer environmental conditions and act accordingly by monitoring the stochastic activity of molecular receptors, whose responses are inherently limited by finite dynamic range \cite{wang1993induction,sourjik2002receptor}, biochemical noise \cite{ten2016fundamental,mora2019physical,tohme2025fast}, and energetic constraints \cite{govern2014energy,liang2024thermodynamic,ouldridge2017thermodynamics,parrondo2015thermodynamics,lestas2010fundamental}. Similar challenges arise in metabolic pathways modulating their propensity to varying nutrient concentrations \cite{solopova2014bet,basan2015overflow}, and adaptive materials in which only a distorted trace of the underlying nonequilibrium dynamics is accessible to engineered sensors \cite{falk2023learning,okumura2022nonlinear}. In all such cases, the system has to tune the strength of its coupling to the signal, whether it is implemented via chemical or mechanical interactions, balancing the information it can extract with the energetic cost of probing \cite{sartori2014thermodynamic,nicoletti2024tuning, nicoletti2025optimal,tjalma2023trade,hartich2015nonequilibrium,skoge2013chemical}. A crucial feature to establish robust functioning is that these operations have to be performed in real time, based solely on data that can be directly observed.

Quantifying this balance from stochastic trajectories proves particularly difficult in realistic scenarios, i.e., when the relevant degrees of freedom (dofs) are only partially observable, and the sensing channel has limited resolution. In cellular signaling, for example, the presence of multiple intermediate processing stages hinders direct observations of receptor binding activity \cite{chang2013adaptive,levchenko2014cellular,fan2024characterizing}, while saturation effects in enzymatic dynamics and chemical noise can alter the sensitivity to ligand concentration \cite{kajita2016balancing,paulsson2000stochastic}. Recent work \cite{nicoletti2024tuning} has shown that approximate estimates of energy consumption during probing can lead to counterintuitive sensing regimes, including cases where an imperfect observer harvests more information than an ideal one, or fails to harvest any information at all. These results underscore the central role of energetics in shaping sensing strategies. Despite providing an interesting proof of concept, the approach presented in \cite{nicoletti2024tuning} remains theoretical, as it relies on force estimation, an inherently challenging task. However, leveraging a recent method to infer dissipation from accessible trajectories using kinetic quantities \cite{di2025force} would open the way to applying these ideas to real biological and synthetic sensing systems.


In this work, we consider a hidden chemical process that stimulates the production of a readout molecule via a transduction channel. A sensor, whether chemical or mechanical, can measure the states of both the readout and the transduction signal, e.g., a receptor binding dynamics or a signaling molecule. The goal of the sensor is to infer, and eventually maximize, the information between the readout and the hidden chemical population, while controlling the overall energetic cost. Crucially, it must do so directly from the trajectories of the dofs it interacts with, relying solely on counting statistics encoded through empirical distributions and finite differences. This framework allows us to characterize how partial observability shapes the optimal balance between information acquisition and dissipation under realistic, real-time physical constraints.

\begin{figure}
    \centering
    \includegraphics[width=\columnwidth]{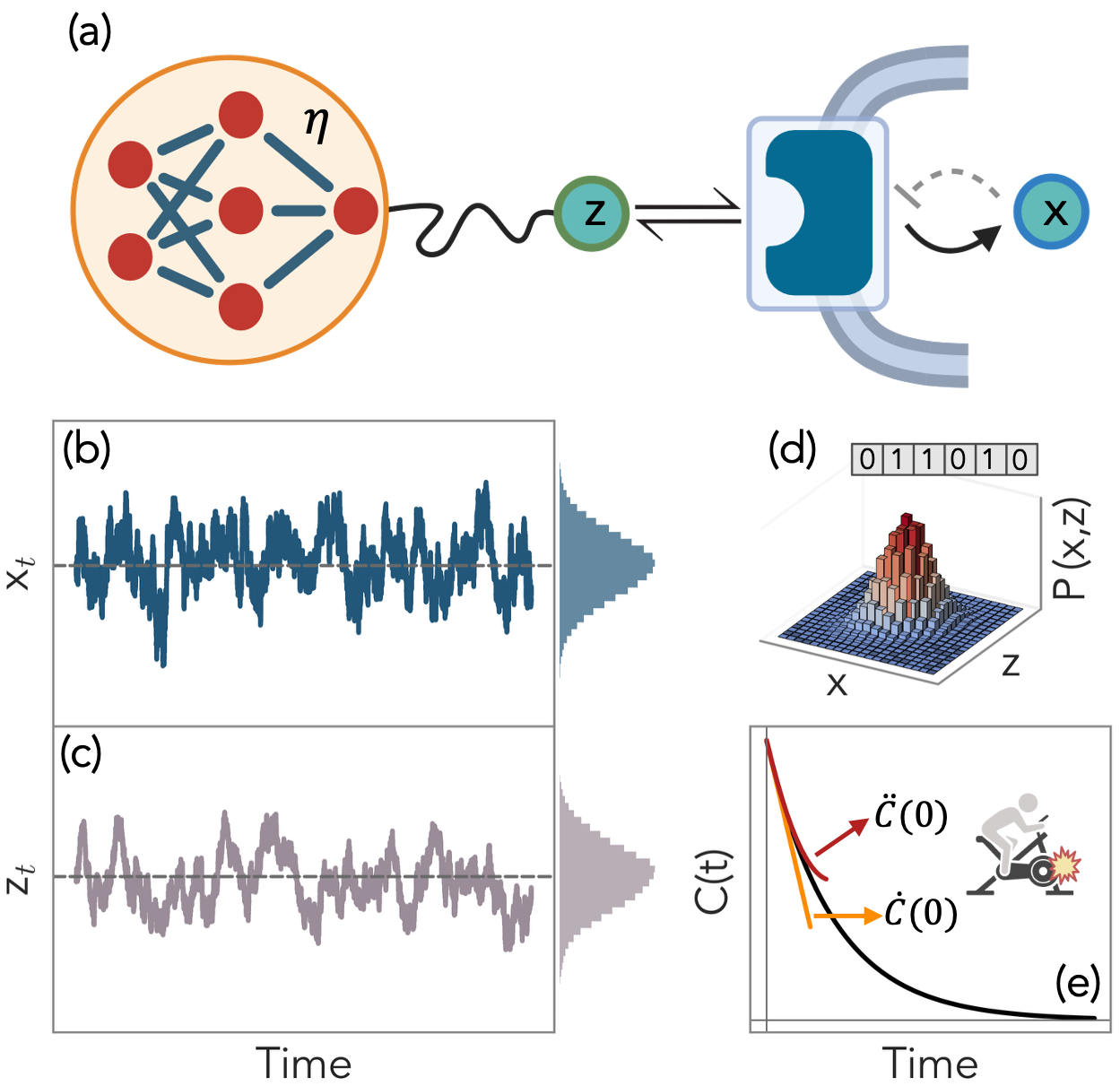}
\caption{(a) Schematic of the sensing architecture. A sensor attempts to gather information about the fluctuations $\vec{\eta}(t)$ of a hidden network, through a readout molecule $X$ coupled to an intermediate signaling molecule $Z$ through receptor binding dynamics. (b-c) Example trajectories of the fluctuations of the intermediate signal $z(t)$ and the corresponding readout $x(t)$. (d) Empirical joint distribution $p^\mathrm{emp}_{xz}$ constructed from the observed trajectories to estimate information (Eq.~\eqref{eqn:pointwise_mutual}). (e) Estimation of the traffic $\mathcal{T}_x$ from the short-time curvature of the autocorrelation function, providing an operational measure of the energetic cost of sensing (Eqs.~\eqref{eqn:traffic_x} and \eqref{eqn:autocorr}).}
\label{fig:figure1}
\end{figure}

We focus on a chemical signaling process, which we sketch in Figure \ref{fig:figure1}a. A hidden chemical network stimulates the production of a signaling molecule $Z$, which in turn binds to a membrane receptor to produce an internal readout $X$. This model can be described by the following set of Langevin equations for concentrations \cite{gardiner2009stochastic}:
\begin{eqnarray}
    \dot{H}_i &=& A^{(H)}_i(\vec{H}) + \vec{b}^{(H)}_i(\vec{H}) \cdot \vec{\xi}_{H} \nonumber \\
    \label{eqn:CLE}
    \dot{Z} &=& A^{(Z)}(\vec{H},Z,X) + b^{(Z)}(\vec{H},Z,X) \xi_Z \\
    \dot{X} &=& A^{(X)}(Z,X) + b^{(X)}(Z,X) \xi_X \nonumber
\end{eqnarray}
where $H_i$ are the concentrations of the hidden molecules. They interact both deterministically via a nonlinear force field with components $A_i^{(H)}$ and through their fluctuations, as $\vec{b}_i^{(H)}$ is a concentration-dependent vector mixing uncorrelated white noises $\vec{\xi}_H$ coming from different hidden reactions. Conversely, $Z$ and $X$ represent the species accessible to the sensor. These dofs only feature one noise source each ($\xi_Z$ and $\xi_X$ with amplitudes $b^{(Z)}$ and $b^{(X)}$) and non-reciprocal deterministic interactions ($A^{(Z)}$ and $A^{(X)}$).
In the End Matter, we show how Eq.~\eqref{eqn:CLE} can be derived from the exemplary chemical model described above by integrating out the (fast) dynamics of intermediate complexes. We are interested in how the sensor reads and processes the fluctuations of concentrations (lowercase letters, $\eta_i$, $z$, and $x$) around the stationary state. Therefore, we employ a linear noise approximation (LNA) \cite{cantini2014linear,grima2015linear}, ending up with:
 \begin{equation}
     \begin{gathered}
        \dotnice{\vec{\eta}} = \tau_\eta^{-1} \vec{W}_\eta \, \vec{\eta} + \sqrt{2 \vec{D}_{\eta}}  \,\vec{\xi}_\eta \\
         \dotnice{z} = - \tau_z^{-1} z + \vec{\sigma} \cdot \vec{\eta} + \alpha\, a\, x + \sqrt{2 D_z} \, \xi_z \\ 
         \dotnice{x} = - \tau_x^{-1} x + a \, g(z) + \sqrt{2 D_x} \, \xi_{x} \\
     \end{gathered}
     \label{eqn:model}
 \end{equation}
where $\vec{\xi}_\eta$, $\xi_z$, and $\xi_x$ are uncorrelated white noises, and $\tau_\eta$, $\tau_z$, and $\tau_x$ characteristic timescales. $D_z$ and $D_x$ are diffusion coefficients that depend in principle on the deterministic fixed points of Eq.~\eqref{eqn:CLE}, and $\vec{D}_{\eta}$ and $\tau_\eta^{-1} \vec{W}_\eta$ are the diffusion and the interaction matrices governing the fluctuations of the hidden dofs. The vector $\vec{\sigma}$ captures the rate of production of the signaling molecule $Z$, while $a$ of the readout $X$. Whenever $X$ is produced, it can feed back on the dynamics of $Z$ through the receptor, e.g., implementing regulatory inhibition mechanisms \cite{cheong2011information,nicoletti2025optimal,ma2009defining,selimkhanov2014accurate,barkai1997robustness,kollmann2005design} (see the gray arrow in Figure \ref{fig:figure1}a). In Eq.~\eqref{eqn:model}, this phenomenon is encoded in $\alpha$, with $\alpha<0$ indicating inhibition. We start by taking $\alpha = 0$, i.e., the signaling process cannot be influenced by its readout, but we will relax this assumption further on. Moreover, on top of the LNA dynamics, we introduce a phenomenological coupling between $Z$ and $X$ as
\begin{equation}
    g(z) = r \tanh\left(\frac{z}{r}\right) \;,
    \label{eqn:tanh}
\end{equation}
where $r$ represents the resolution with which the readout can resolve the signal. If $r \to \infty$, $g$ becomes a linear function, whereas for small $r$, $X$ can only detect small fluctuations. This finite resolution, while approximate, may be relevant in more general settings, such as chemical systems with kinetic traps or catalytic couplings, and models of gene activity \cite{inoue2013cooperative,pham2024dynamical,Olmeda2025.07.09.663203}.

\begin{figure*}
    \centering
    \includegraphics[width=\textwidth]{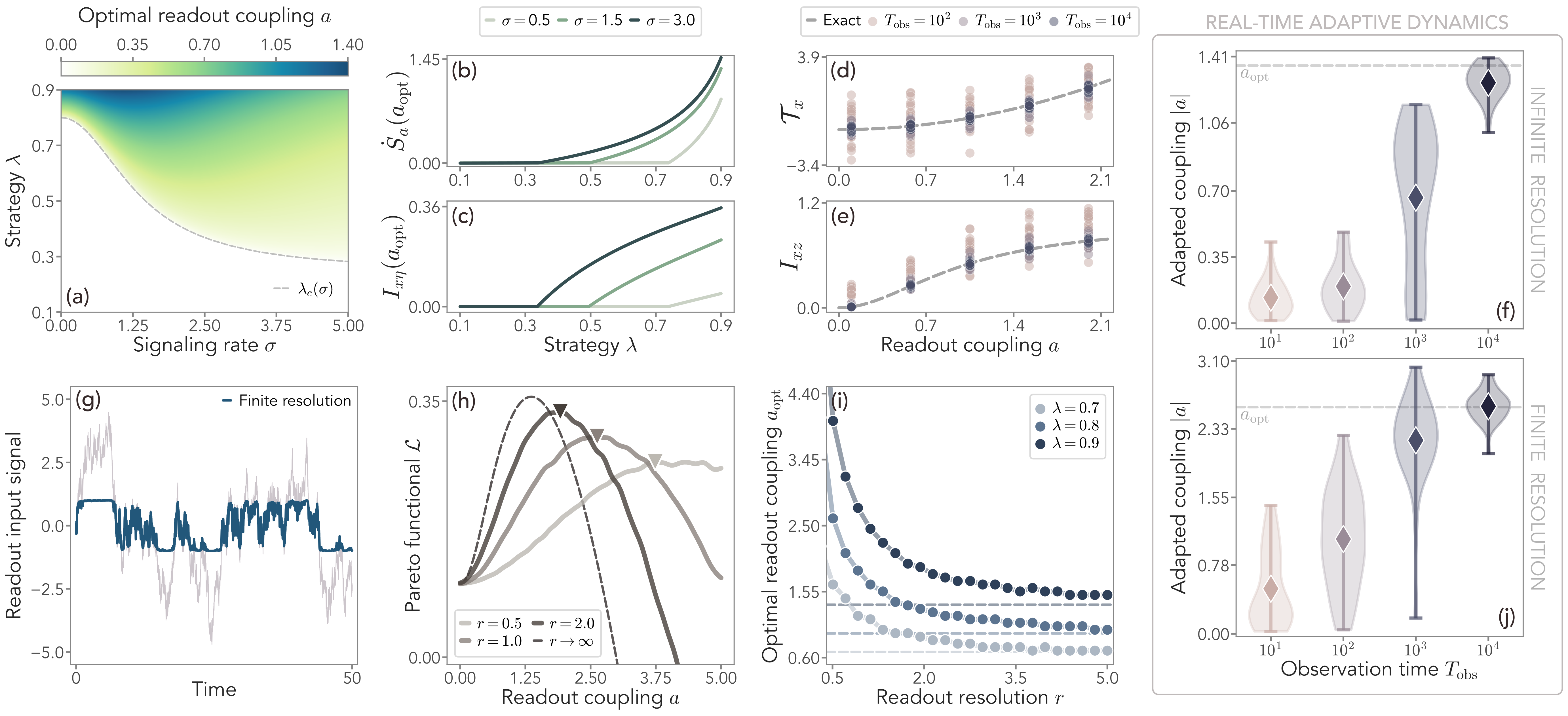}
    \caption{(a) Optimal readout coupling $a_\mathrm{opt}$ for infinite resolution as a function of the strategy $\lambda$ and signaling rate $\sigma$. At fixed $\sigma$, we find a transition from a regime where $a_\mathrm{opt}$ vanishes at low $\lambda$ and increases for $\lambda > \lambda_c$ (grey dashed line). (b-c) Entropy production, $\dotnice{S}_a$, and information between $x$ and $\eta$, $I_{x\eta}$, as a function of the strategy $\lambda$ for different values of $\sigma$, when $a = a_\mathrm{opt}$. (d) Estimate of $\mathcal{T}_x$ from trajectories of finite duration $T_\mathrm{obs}$. As $T_\mathrm{obs}$ increases, the estimated value approaches the theoretical one (gray dashed line). (e) Same, but for the information between the readout and the signal, $I_{xz}$. (f) Adapted readout coupling after $10^4$ steps of adaptive dynamics, where the sensor attempts to tune $a$ by estimating $\mathcal{T}_x$ and $I_{xz}$ from measurement windows of duration $T_\mathrm{obs}$. For large enough $T_\mathrm{obs}$, the readout achieves a near-optimal coupling. Distributions are obtained over $64$ repetitions of the adaptive dynamics. (g) Example of the readout input signal, both for infinite resolution (gray trajectory) and for a resolution $r = 1$ (blue trajectory). (h) Comparison of the functional $\mathcal{L}$ at different resolution values. The maximum (triangles) moves at higher values of $a$ as $r$ decreases. (i) Optimal readout coupling as a function of the resolution. As the resolution increases, $a_\mathrm{opt}$ decreases until it reaches the $r\to\infty$ limit (dashed lines). (j) Same as (f), but for a finite resolution ($r=1$). In this figure, unless otherwise specified, $D_i = 1/\tau_i$ for $i=\eta,z,x$, $W_\eta = -1$, $\sigma = 1.5$, $\lambda = 0.9$, $\tau_\eta = 3\tau_z$, $\tau_z = \tau_x = 1$.}
    \label{fig:figure2}
\end{figure*}

The sensor has no direct access to the statistics of $\vec{\eta}$. Rather, at any time $t$, it can only measure the fluctuations of the readout, $x(t)$, and that of the signal $z(t)$ (Figure \ref{fig:figure1}b-c). 
An enzyme converting $X$ into a molecule triggering allosteric changes in the membrane receptor might be a concrete example of such a sensor, which can record $X$ and $Z$ through direct couplings. Furthermore, this implies that the sensor can tune the coupling $a$ by accelerating or slowing down the production of $X$ via allosteric interactions \cite{flatt2023abc}. Thus, from these limited observations, the sensor can set $a$ in such a way that it gathers as much information as possible on $\vec{\eta}$, while at the same time balancing this information gain with the energy it dissipates to achieve it. The sensor can obtain a fluctuating measure of the information between $x(t)$ and $z(t)$ by the pointwise mutual information \cite{parrondo2015thermodynamics,nicoletti2024tuning}:
\begin{equation}
\label{eqn:pointwise_mutual}
    i_{xz}(t) = \log\frac{p^\mathrm{emp}_{xz}(x(t), z(t))}{p^\mathrm{emp}_x(x(t)) \, p^\mathrm{emp}_z(z(t))}
\end{equation}
with $p^\mathrm{emp}_{xz}$ the joint empirical distribution (Figure \ref{fig:figure1}d), and $p^\mathrm{emp}_{x}$ and $p^\mathrm{emp}_{z}$ the marginal ones. The mutual information between $x$ and $z$ is the average of Eq.~\eqref{eqn:pointwise_mutual} over the observable times, i.e., $I_{xz} = \ev{i_{xz}}$. We remark that there is no guarantee apriori that increasing $I_{xz}$ will also increase the mutual information between $x$ and $\vec{\eta}$, $I_{x\vec{\eta}}$. Nevertheless, as we will see, this is often an effective strategy that allows the sensor to tune its allosteric modulations so that the readout reflects the activity of $\vec{\eta}$.

Being non-reciprocal, the coupling $a$ leads to unavoidable dissipation \cite{Agudo_Active_Phase,loos2020irreversibility,suchanek2023entropy,suchanek2023entropy,Zhang_EP_2023,dinelli2023non,busiello2024unraveling,pham2025irreversibility}. A direct evaluation of the associated entropy production $\dotnice{S}$ is highly challenging, as it would require, in principle, estimating the force field between $X$ and $Z$ from stochastic trajectories \cite{sekimoto1998langevin,frishman2020learning}. However, the traffic $\mathcal{T}_x$ \cite{maes2020frenesy,di2018kinetic,MAES20201} provides a more accessible alternative: it can be extracted directly from the short-time fluctuations of the readout $X$, relying solely on its counting statistics (see End Matter). For the x component, it is defined as
\begin{equation}
\label{eqn:traffic_x}
    4 \mathcal{T}_x = \frac{\ddotnice{C}_{xx}(t = 0)}{\dotnice{C}_{xx}(t = 0)} \;,
\end{equation}
where $C_{xx}(t)$ is the autocorrelation function,
\begin{equation}
\label{eqn:autocorr}
    C_{xx}(t) = \ev{x(t) x(0)} - \ev{x}^2 \; ,
\end{equation}
and serves as a proxy for the entropy production of the system \cite{maes2008steady,di2025force} (see Figure \ref{fig:figure1}e and End Matter).

Having measured both information and dissipation from observed fluctuating trajectories, the sensor can implement, autonomously or by design, a strategy $\lambda$ to balance them. Following previous work \cite{nicoletti2024tuning}, we introduce the Pareto functional \cite{shoval2012evolutionary},
\begin{equation}
    \mathcal{L}(a, \lambda) =  \lambda \ev{i_{xz}}(a) - 4 (1-\lambda) \mathcal{T}_x(a)
    \label{eqn:functional_generic}
\end{equation}
where $\lambda \in [0,1]$, and both $I_{xz}$ and $\mathcal{T}_x$ depend on the coupling strength $a$ (see Eq.~\eqref{eqn:tanh}). At a given strategy $\lambda$, we set the value of $a$ that maximizes information while keeping dissipation as low as possible, i.e., $a_\mathrm{opt}(\lambda) = \text{argmax}_a \mathcal{L}(a, \lambda)$. For small $\lambda$, the chemical sensor acts to preferentially minimize the estimated dissipation, whereas large $\lambda$ denotes an information-driven strategy that may incur a higher energetic cost. We stress that the functional in Eq.~\eqref{eqn:functional_generic} only contains quantities that are measurable by the sensor. 

We first focus on the ideal case of an infinite resolution $r\to\infty$, where the sensor can measure the entire dynamic range of $z$ for an infinite observation window, $T_\mathrm{obs} \to \infty$. In this case, Eq.~\eqref{eqn:model} 
can be solved exactly, so that $\mathcal{L}(a)$ can be obtained analytically. Moreover, for simplicity, we set $\vec{\eta}$ to be one-dimensional, but the proposed approach can be extended to high-dimensional inaccessible dofs. In Figure \ref{fig:figure2}a, we show that $a_\mathrm{opt}(\lambda)$ displays a sharp transition at $\lambda = \lambda_c(\sigma)$, at all values of $\sigma$. Indeed, when the strategy is dissipation-driven (small $\lambda$), optimal balance prescribes that no information is harvested, and $a_\mathrm{opt}$ vanishes. However, as $\lambda$ increases, the strategy becomes more information-driven, and the optimal coupling eventually becomes non-zero and grows accordingly. Crucially, increasing the strategy $\lambda$ over a critical value $\lambda_c$ also corresponds to an increase in the information between the readout $x(t)$ and the underlying hidden process $\eta(t)$, $I_{x\eta}$, while keeping the entropy production induced by the readout coupling, $\dotnice{S}_a = \dotnice{S} - \dotnice{S}|_{a=0}$, as low a possible (see Figure \ref{fig:figure2}b-c). This shows that the chemical sensor can construct - and ultimately optimize - efficient representations of information about the inaccessible process and dissipation using only empirical statistics on observed quantities.

When $T_\mathrm{obs}$ is finite, however, both Eq.~\eqref{eqn:pointwise_mutual} and Eq.~\eqref{eqn:autocorr} suffer from sampling errors. Importantly, this closely resembles the common situation of a chemical sensor that must act on the fly. To model this situation, we now consider the case of a sensor that continuously changes $a$ for $N$ adaptation steps of duration $T_\mathrm{obs}$ each. During the $n$-th step, the sensor sets $a = a_n$ and constructs an estimate of the quantities in Eq.~\eqref{eqn:functional_generic} by observing trajectories for the finite time window. Then, the sensor changes the coupling to $\tilde{a}_{n+1} = a_n + \Delta a$, with $\Delta a \sim \mathcal{N}(0,\sigma_a)$ and performs a new trajectory-based estimation of Eq.~\eqref{eqn:pointwise_mutual} and Eq.~\eqref{eqn:traffic_x}. Based on this temporally-constrained observations, the new coupling is accepted ($a_{n+1} = \tilde{a}_{n+1}$) if $\mathcal{L}(\tilde{a}_{n+1}) > \mathcal{L}(a_{n})$, otherwise the coupling goes back to the previous value ($a_{n+1} = a_n$). This adaptive dynamics proceeds until no new adaptive move is accepted for a sufficiently long stopping time. Notice that we start from a perfectly inhibiting sensor that impedes $x(t)$ from having any information on $\vec{\eta}(t)$, $a_0 = 0$. First, in Figure \ref{fig:figure2}d-e, we show that, if $T_\mathrm{obs}$ is particularly small, sampling errors may lead to a large variance in the estimation of both $\mathcal{T}_x$ and $I_{xz}$, especially at large $a$. Then, in Figure \ref{fig:figure2}f we plot the distribution of $a$ at convergence of the adaptive dynamics as a function of $T_\mathrm{obs}$. Remarkably, as $T_\mathrm{obs}$ increases, the average of the distribution approaches $a_\mathrm{opt}$, allowing the sensor to reach the optimal coupling through purely trajectory-based estimates.

\begin{figure}[t]
    \centering
    \includegraphics[width=\columnwidth]{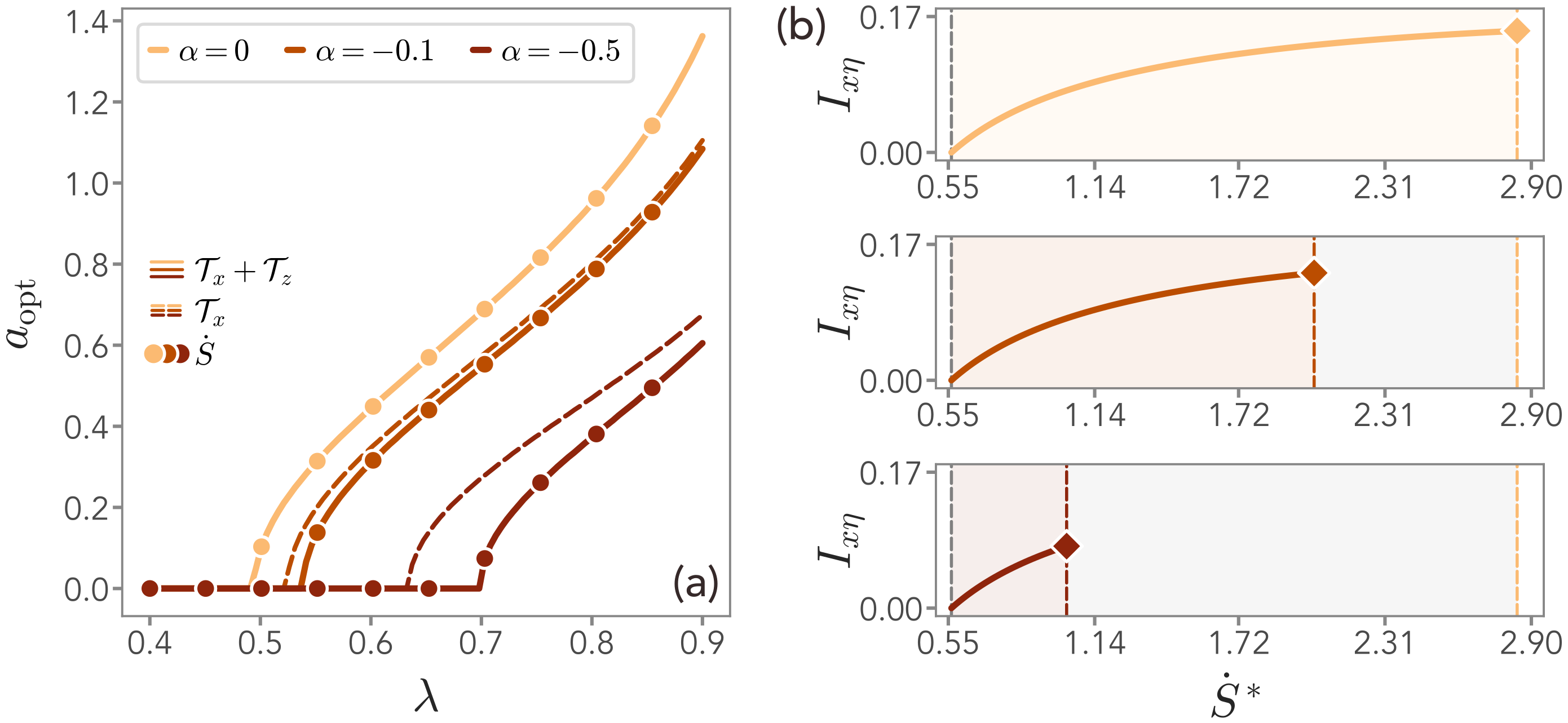}
    \caption{(a) Optimal readout coupling $a_\mathrm{opt}$ as a function of the strategy $\lambda$. Without an inhibitory feedback coupling ($\alpha = 0$), it is enough to estimate the $x$-component of the traffic $\mathcal{T}_x$, which contains all relevant dependences of $\dotnice{S}$ on $a$. When $\alpha < 0$, both $\mathcal{T}_x$ and $\mathcal{T}_z$ are needed, since the functional with the $x$-component alone overestimates $a_\mathrm{opt}$. As before, there is a critical $\lambda_c$ below which $a_\mathrm{opt} = 0$. (b) Information between readout and hidden dof, $I_{x\eta}$, as a function of the energy consumption $\dotnice{S}^*$ of the system, which is the entropy production at the corresponding optimal coupling $a_\mathrm{opt}$. Here, $\lambda \in [0.1, 0.9]$. Since $a_\mathrm{opt}$ decreases with more negative values of $\alpha$, so does the energy consumption. Thus, the maximum value of $\dotnice{S}^*$ (squares) is larger for $\alpha = 0$. The gray dashed line represents the baseline energy consumption when the readout does not interact with the signal, for $a = 0$. In this figure, $D_i = 1/\tau_i$ for $i=\eta,z,x$, $W_\eta = -1$, $\sigma = 1.5$, $\tau_\eta = 3 \tau_z$, $\tau_z = \tau_x = 1$.}
    \label{fig:figure3}
\end{figure}

An important ingredient in most signaling networks is the presence of regulatory inhibition mechanisms \cite{cheong2011information,nicoletti2025optimal,ma2009defining,selimkhanov2014accurate,barkai1997robustness,kollmann2005design} coupling the readout $X$ with the signaling molecule $Z$, encoded in the parameter $\alpha<0$ in Eq.~\eqref{eqn:model}. For simplicity, we stick to the $r \to \infty$ limit as it does not qualitatively change the results. Because of this feedback coupling, the sensor must now estimate dissipation by evaluating the traffic for both the observable dofs, $\mathcal{T}_x$ and $\mathcal{T}_z$, defined by their autocorrelation functions (see Eq.~\eqref{eqn:traffic_x} and End Matter). Since the sensor has access to the trajectories of both the readout and the signal, this remains a feasible estimation with counting statistics alone. In Figure \ref{fig:figure3}a, we plot the optimal readout coupling as a function of the strategy $\lambda$ for different values of $\alpha$. When $\mathcal{T}_x + \mathcal{T}_z$ is used by the sensor, the optimal coupling matches the one obtained in the ideal case where the entire entropy production $\dot{S}$ is known. When only $\mathcal{T}_x$ is employed and $\alpha \ne 0$, on the other hand, $a_\mathrm{opt}$ becomes systematically larger due to the fact that the sensor underestimates the dissipation. Furthermore, as $\alpha$ decreases, the strategy needs to be more and more information-driven for $a_\mathrm{opt}$ to be different from zero. This suggests that the regulatory mechanism is effectively increasing the cost of acquiring information, pushing $\lambda_c$ to higher values.

Finally, we set a given energy budget -- i.e., we fix the total energy consumption of the system $\dotnice{S} = \dotnice{S}^*$, which in turn determines the strategy of the sensor -- and assess the performance of the optimal coupling by evaluating the information between the readout and the hidden process, $I_{x\eta}$. We recall that, although $I_{x\eta}$ cannot be measured directly, it is the ideal target that the sensor wants to maximize. In Figure \ref{fig:figure3}b, we plot $I_{x\eta}$ against $\dotnice{S}^*$ up to the value corresponding to $\lambda = 0.9$. When $\alpha = 0$, there is no regulatory mechanism in place and the sensor can efficiently use energy to harvest information. However, as $\alpha$ decreases, the maximum value of $I_{x\eta}$ achievable at increasing energy budget becomes smaller for the same set of strategies. Remarkably, this behavior is still optimal in terms of the limited observability of the system -- that is, the strategy being equal, the sensor tunes the readout coupling to account for the increased cost of information and aims at reducing $\dotnice{S}^*$ accordingly.

Overall, we have shown that trajectory-based estimates relying on counting statistics can be successfully implemented using pointwise mutual information and dissipation estimation via the traffic. This is particularly relevant when not all of the degrees of freedom can be observed, as in the case of the chemical sensor studied here. Furthermore, such estimates can be exploited to balance information and dissipation during sensing processes. Despite our focus on a chemical network, this conceptual framework is in principle applicable to any artificial or biological system operating under similar constraints, especially multiscale ones \cite{nicoletti2024information, nicoletti2024gaussian, paulsson2000stochastic} or coupled to shared changing environments \cite{nicoletti2021mutual, karin2021temporal, mora2019physical, malaguti2021theory}. We remark here that information maximization and dissipation minimization may not be general driving principles, as these two quantities are not necessarily in tradeoff \cite{baiesi2018life}. Nevertheless, several studies have shown that they appear in different contexts, either separately or together, from sensory adaptation \cite{nicoletti2025optimal, lan2012energy} to bacterial chemotaxis \cite{mattingly2021escherichia,hathcock2023nonequilibrium} and chemical networks \cite{flatt2023abc, govern2014energy}. Our approach outlines a concrete path toward testing the relevance of these principles in biological systems, where the balance between energy consumption and information gathering can be incorporated as a plausible estimate from observed stochastic trajectories. In future works, it will be interesting to replace the mutual information with other quantities that can be measured in the presence of high-dimensional signals, where reliably estimating empirical probabilities becomes increasingly harder. Moreover, the application of these ideas to detailed signaling networks is a fascinating possibility that may lead to uncovering the specific information-theoretic and thermodynamic role of known biochemical couplings. We believe that this study provides explicit evidence that, under physically relevant constraints, balancing information and dissipation might be an implementable design principle to perform adaptive strategies in biological and biochemical systems.

\begin{acknowledgments}
\noindent The authors thank Marco Baiesi, Matteo Ciarchi, and Vincenzo Maria Schimmenti for useful discussions. The authors thank the Max Planck Institute for the Physics of Complex Systems for hosting G.N. and D.M.B. during the initial phase of this work, and the International Center for Theoretical Physics for hosting D.M.B. during the final phase of this work. D.M.B. is funded by the program STARS@UNIPD with the project ``ActiveInfo''.
\end{acknowledgments}









\bibliography{Bibliography}

\onecolumngrid
\begin{center}
\vspace{0.5cm}
\textbf{\large End Matter}\\
\end{center}
\twocolumngrid

\noindent\textit{An exemplary signaling model} -- Here, we introduce a chemical network that plays the role of an elementary example of the class of systems studied in this work. The internal dynamics producing a signaling molecule $Z$ is:
\begin{gather*}
    \sum_j \alpha_j H_j \rightleftharpoons \sum_j \beta_j H_j \nonumber \\ 
    \sum_j \gamma_j H_j \, + \, E \rightleftharpoons C_E \rightarrow E \, + \, Z \;,
\end{gather*}
where we maintain full generality for the internal dynamics of $H_i$ species, and we propose a simple, yet general, form of the process producing $Z$, i.e., an enzyme $E$ catalyzing the conversion of internal molecules through an intermediate complex $C_E$. Then, $Z$ can interact with a membrane receptor $R$ to generate $X$ via the complex $C_R$:
\begin{equation*}
    Z \, + \, R \rightleftharpoons C_R \rightarrow R + X \qquad X \rightarrow \varnothing \;,
\end{equation*}
where the rightmost reaction represents a degradation dynamics. The associated stochastic evolution for concentrations (upper-case letters) can be obtained via Kramers-Moyal expansion \cite{gardiner2009stochastic} and feature non-linear reaction terms and non-diagonal multiplicative noises.

\begin{figure}
    \centering
    \includegraphics[width=0.9\columnwidth]{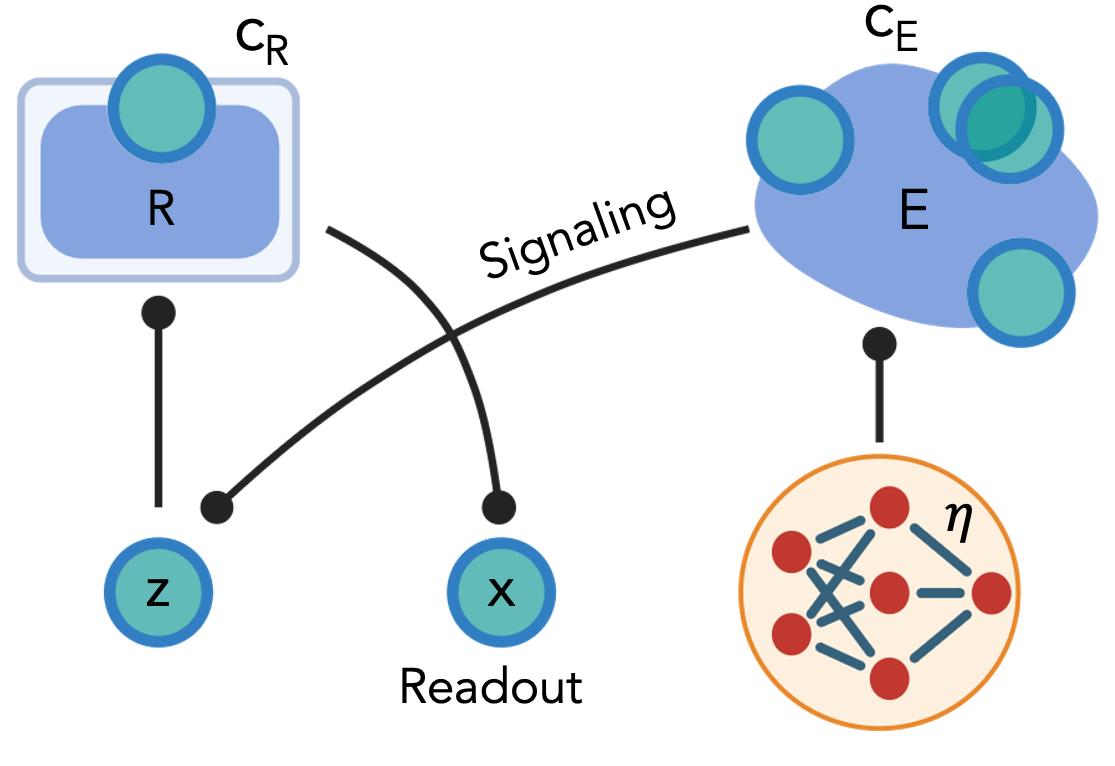}
\caption{Scheme of the chemical model. Internal species $\vec{H}$, readout $X$, and signaling molecule $Z$ (whose fluctuations are in lowercase letters, $\vec{\eta}(t)$, $x(t)$, and $z(t)$) are coupled together only through intermediate fast complexes.}
\label{fig:figureE1}
\end{figure}

We first incorporate the conservation laws $R(t) = R_{\rm tot} - C_R(t)$ and $E(t) = E_{\rm tot} - C_E$. Since typically complexes are fast variables, as for nucleotide-exchange and several enzyme kinetics \cite{flatt2023abc,liang2024thermodynamic}, the chemical system has a layered structure as in Figure \ref{fig:figureE1}. Since all couplings between $\vec{H}$, $X$, and $Z$ are mediated by fast species, by writing down the Fokker-Planck equation, and integrating out $C_R$ and $C_E$ (with no-flux boundary conditions for all species), we obtain a new dynamics that only features diagonal multiplicative noise terms in $Z$ and $X$. Indeed, there are no reactions (hence no noise) directly mixing $\vec{H}$, $Z$, and $X$ (see Figure \ref{fig:figureE1}). Moreover, the drift terms will depend on the average of $C_E$ and $C_R$ that, in the timescale separation regime, only depend on $\vec{\eta}$ and $Z$, respectively. The structure of the resulting dynamics resembles the one presented in Eq.~\eqref{eqn:CLE}. We remark here that we are only interested in motivating the use of our minimal model, without drawing any detailed connection between microscopic and effective parameters, since this will depend on the entire chemical dynamics.

The presence of a negative feedback from $X$ on $Z$ can also be implemented through a reaction:
\begin{equation}
    X + R \rightleftharpoons C_R \;.
\end{equation}
where the additional (fast) complex formation effectively modifies the concentration of $Z$ when integrated out.

For simplicity, limited resolution is added on top of the linearized dynamics to capture the ability to resolve only a certain range of fluctuations. While phenomenological at this stage, a similar threshold dynamics emerges in the microscopic model as a consequence of the saturating behavior of $C_R$ as a function of $z$.

\noindent\textit{Entropy production in diffusive systems and kinetic inference}—
Consider a general multidimensional overdamped Langevin dynamics
\begin{equation}
    \dot{\vec{x}} = \vec{A}(\vec{x})
    + \sqrt{2\vec{D}}\,\vec{\xi},
    \qquad
    \langle \xi_i(t)\,\xi_j(s) \rangle
    = \delta_{ij}\delta(t-s),
\end{equation}
with drift $\vec{A}(\vec{x})$ and diffusion matrix $\vec{D}$.  
The associated Fokker–Planck equation reads
\begin{equation}\label{FP_supp}
    \partial_t p(\vec{x})
    = -\nabla\!\cdot\!\vec{j}(\vec{x}),
\end{equation}
and in a nonequilibrium steady state (NESS) one has $\partial_t p(\vec{x})=0$ and a nonvanishing solenoidal probability current $\vec{j}(\vec{x}) = \left[\vec{A}(\vec{x}) - \nabla\phi(\vec{x})\right] p(\vec{x}) \neq 0$, $\nabla\!\cdot\!\vec{j}(\vec{x})=0$. The NESS is characterized by a constant rate of entropy production $\dot{S}$ \cite{maes03_1,seifert2012stochastic,peliti2021stochastic,GalCohenFluc,fluc1}, which can be identified with the heat dissipated into the environment as prescribed by stochastic energetics~\cite{sekimoto1998langevin}:
\begin{equation}\label{sekimoto}
    \dot{S}
    = \langle \vec{D}^{-1}\vec{A}(\vec{x})
      \circ \dot{\vec{x}}\,\rangle.
\end{equation}
Estimating $\vec{A}(\vec{x})$ directly from trajectories is often impractical, motivating alternative approaches.  
A particularly useful one is a kinetic decomposition of the entropy production~\cite{maes2008steady,frishman2020learning,di2025force},
\begin{equation}\label{eq:general_EP}
    \dot{S} = 4\,\mathcal{T} + \inflow,
\end{equation}
which expresses $\dot{S}$ in terms of two quantities that can, in principle, be inferred from time-series data.  
The \emph{traffic} $\mathcal{T}$ quantifies the time-symmetric dynamical activity, whereas the \emph{inflow rate} $\inflow$~\cite{fal15c} captures the local compression or expansion of probability. Using a recently introduced variance sum rule~\cite{VSR,di2025force}, the traffic can be written as
\begin{equation}\label{eq:traffic_general}
    \mathcal{T}
    = -\frac{1}{4}\,(D^{-1})_{ij}\,
      \ddot{C}^{(x)}_{\,ij}(0),
\end{equation}
where $C^{(x)}_{\,ij}(t) = \langle x_i(t)\,x_j(0)\rangle - \langle x_i\rangle\langle x_j\rangle$
is the autocorrelation matrix of the observed trajectory $\vec{x}(t)$.  
The diffusion coefficients entering Eq.~\eqref{eq:traffic_general} are also accessible from short-time statistics: $D_{ij} = -\dot{C}^{(x)}_{(ij)}(0)$, with $
C_{(ij)} = (C_{ij}+C_{ji})/2$. On the other hand, the inflow rate can be both expressed in terms of $\nabla\phi = -\nabla \log p(x_t)$, which can be estimated from trajectories and in terms of the divergence of the drift,
\begin{equation}\label{eq:inflow_general}
    \inflow
    = D_{ij}\,C_{\, ij}^{\nabla\phi}(0) = -\int d\vec{x}\;
      p(\vec{x})\,\partial_i A_i(\vec{x}) \;.
\end{equation}
Equations~\eqref{eq:general_EP}-\eqref{eq:inflow_general} show that both contributions to $\dotnice{S}$ can be expressed solely in terms of short-time correlations of observable dynamical variables.

In our model capturing fluctuations, Eq.~\eqref{eqn:model}, the diffusion matrix is diagonal, $D_{ij}=D_i\delta_{ij}$ in the observable quantities, and the sensor modifies only the coupling between the signaling molecule $Z$ and the readout $X$. Therefore, since we are interested in estimating the entropy production induced by the non-reciprocity of $a$,
\begin{eqnarray}
    \dotnice{S}_a = \dotnice{S} - \dotnice{S}|_{a = 0} \;,
\end{eqnarray}
the only surviving terms are those affected by a modification in the value of the coupling $a$. As such, the inflow rate is independent of $a$ and the relevant terms in the traffic are $\mathcal{T}_x$ and $\mathcal{T}_z$, defined as $-\frac{1}{4}D_i^{-1}\,\ddot{C}_{\,ii}^{(\beta)}(0)$ with $\beta = x,z$. Consequently,
\begin{equation}
    \dotnice{S}_a = 4\, \left[\left(\mathcal{T}_x + \mathcal{T}_z\right) - \left(\mathcal{T}_x + \mathcal{T}_z\right)|_{a = 0} \right] \;.
\end{equation}
Notice that, when $\alpha = 0$, the only remaining term is $\mathcal{T}_x$, since $\mathcal{T}_z$ would not depend on $a$. 

Numerically, we estimate the autocorrelation functions and their derivatives directly through finite differences. Since the traffic depends only on short-time statistics of the accessible variables, it is particularly suited for inferring changes in dissipation in partially observed systems as the one studied in this work, where force fields cannot be easily reconstructed.

\noindent\textit{Exact results in the infinite resolution limit} -- In the infinite resolution limit, Eq.~\eqref{eqn:model} reduces to an Ornstein-Uhlenbeck process with interaction matrix
\begin{equation}
    \vec{A} = \begin{pmatrix}
        \vec{W}_\eta / \tau_\eta & 0 & 0 \\
        \vec{\sigma}  & -1/\tau_z & \alpha\, a  \\
        0 & a & -1/\tau_x
    \end{pmatrix}
\end{equation}
and a block-diagonal diffusion matrix $\vec{D} = \text{diag}(\vec{D}_{\eta} , D_z, D_x)$. The covariance matrix $\vec{\Sigma}$ of the entire system obeys the Lyapnuov equation $\vec{A}\vec\Sigma + \vec\Sigma\vec{A}^T = -2\vec{D}$ and can be found analytically. Then, if $T_\mathrm{obs}$ is also large and we can neglect sampling errors, the components of traffic estimated from the stochastic trajectories become exact and are given by the diagonal elements of the matrix $\vec{\Xi} = \vec{D}^{-1}\vec{A}\vec{\Sigma}\vec{A}^T - \vec{A}$. In particular, in the main text we considered $\mathcal{T}_x = \Xi_{xx}$ and $\mathcal{T}_z = \Xi_{zz}$, whereas the entropy production if $\dot{S} = \Tr\,\vec{\Xi}$. In this limit, on the other hand, the average of the pointwise mutual information in the $T_\mathrm{obs}\to\infty$ limit becomes
\begin{equation}
    I_{xz} = \frac{1}{2}\log\frac{\Sigma_{xx}\Sigma_{zz}}{\Sigma_{xx}\Sigma_{zz}-\Sigma_{xz}^2}
\end{equation}
and similarly for $I_{x\vec{\eta}}$ \cite{nicoletti2024tuning}.





\end{document}